\documentstyle[epsf,pjournal,mamd,natbib]{aa}

\begin{document}

\thesaurus{23(03.09.1, 03.20.1, 13.09.2)}
\title{Optimizing ISOCAM data processing using spatial redundancy
\thanks{Based on observations with ISO, an ESA project with instruments
funded by ESA Member States (especially the PI countries: France, Germany,
the Netherlands and the United Kingdom) and with the participation of ISAS
and NASA}}

 \author{M.-A. Miville-Desch\^enes\inst{1}\inst{2} \and F. Boulanger\inst{1} \and A. Abergel\inst{1} \and 
 J.-P. Bernard\inst{1}}
 \institute{Institut d'Astrophysique Spatiale, Universit\'e Paris-Sud, B\^at. 121, 91405, Orsay, France
  \and D\'epartement de Physique, Observatoire Astronomique du Mont M\'egantic, Universit\'e Laval, 
 Sainte-Foy, Qu\'ebec, G1K 7P4, Canada}

 \offprints{Marc-Antoine Miville-Desch\^enes}
 \mail{mamd@ias.fr}
 \date{Received April 16, 1999 / Accepted June 7, 1999}
% \date{\today}

\titlerunning{ISOCAM processing using spatial redundancy}
 \maketitle

\begin{abstract}

Several instrumental effects of the Long Wavelength channel of ISOCAM, the camera
on board the Infrared Space Observatory, degrade the processed images.
We present new data processing techniques that correct these defects, 
taking advantage of the fact that a position on the sky has been observed by several 
pixels at different times. We use this redundant information (1) to correct the 
long term variation of the detector response, 
(2) to correct memory effects after glitches and point sources, and (3) to refine the deglitching process.
As an example we have applied our processing on the gamma-ray burst observation GRB 970402.
Our new data processing techniques allow the detection of faint extended emission with contrast
smaller than 1\% of the zodiacal background.
The data reduction corrects instrumental effects to the point where
the noise in the final map is dominated by the readout and the photon noises.
All raster ISOCAM observations can benefit from the
data processing described here. This includes mapping of solar system extended objects (comet dust trails), 
nearby clouds and star forming regions, images from diffuse emission in the Galactic plane
and external galaxies.
These techniques could also be applied to other raster type observations (e.g. ISOPHOT).

\end{abstract}

\section{Introduction}

The Infrared Space Observatory (ISO)
 of the European Space Agency died on
April 8 1998, 10 months after its expected end. This mission has been
a great success \cite[]{kessler99} and many of the scientific discoveries are yet to come.
Much of the scientifical analysis of the data are still held by data reduction problems.
In particular for the camera on board of ISO (ISOCAM - \cite{cesarsky96}), 
instrumental effects which are not well understood limit the sensitivity.

Much effort has been made to model the response of the ISOCAM array based on theoretical
understanding of infrared detectors (\cite{abergel99}, \cite{coulais2000} and references therein).
Nevertheless, no overall model of the ISOCAM response to flux steps and glitches is available.
ISOCAM data reduction thus relies on an empirical understanding of the 
detector. Recently, sophisticated data reduction techniques have been developped
to take into account some aspects of ISOCAM response 
(\cite{starck99}, \cite{desert99}, \cite{aussel99},  
\cite{altieri99}). These methods are close to being optimal for the detection of point sources. 
Nevertheless, in many observations, instrumental effects are still preventing 
the study of faint extended emission.

The development of the data reduction method presented in this paper was motivated by the analysis of 
ISOCAM observations of diffuse and translucent interstellar clouds.
Such clouds, when illuminated by the solar neighborhood radiation field
have mid-infrared brightness with very low contrast with respect to the zodiacal light (at most
a few percent).
To unveil these low signal-to-noise ratio clouds, variations in the detector response
have to be corrected to an accuracy better than a fraction of 1\%.
At the present time, data processing techniques available do not allow 
to achieve such high sensitivity for extended emission.
In this paper we present a data processing method that make it possible.

The algorithms presented here apply to observations made in raster mode with
the Long Wavelength channel of ISOCAM (LW). It makes use of the fact that a given
position on the sky has been observed several times by different camera pixels.
Tests have been performed on extended emission observations presenting low and high contrasts. We
have checked that our methods 
can be applied to most raster mode observations, and as a consequence concerns 
a significant fraction of ISOCAM observations (extragalactic, galactic and even solar system observations). 

To illustrate the data processing we use two different observations of the same field, 
obtained subsequently in exactly the same configuration. The amplitude of instrumental 
effects are not the same in both observations, giving us constraints on the validity of our method.
In section \S~\ref{observation} we present the observations used to 
illustrate the data reduction chain.
In section \S~\ref{std_data_processing} standard data reduction techniques
are briefly presented. The main problems of the standard reduction are described in \S~\ref{problems}
and the new data processing approach to address these problems is detailed in \S~\ref{new}.
The performances of the overall method are discussed in \S~\ref{discussion}.

\section{Observation}

\label{observation}

The main ISOCAM technical characteristics are presented in \cite{cesarsky96}. 
The LW channel of ISOCAM operates between 4 and 18 $\mu$m. A lens wheel allows the selection of the 
field of view per pixel (1.5, 3, 6 and 12 arcsecs), and 
a filter wheel the selection of the spectral band pass (10 broad band
filters and two Continuously Variable Filters). The detector is a 500~$\mu$m thick crystal
made of Gallium doped Silicon photo-conductor hybridized by Indium bumps. 
The 32$\times$32 square pixels are defined with a pitch of 100~$\mu$m.

\begin{figure}
\epsfxsize=9cm
\epsfbox{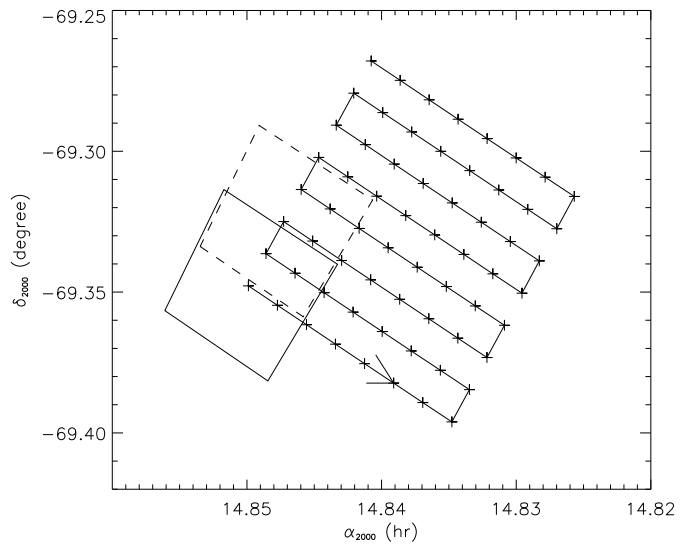}
\caption{\label{strategie} Raster observing strategy. 
The crosses indicate the positions on the sky of the camera center.
The first (solid line) and the sixteenth (dash line)
 positions of the camera are shown as {\it square}. The arrow shows the scanning direction.}
\end{figure}

A given observation (characterized by a given configuration (lens, filter, integration time))
is presented as a collection 
of $32\times32$ images (called readouts) gathered together in a data cube.
Three different observational modes were available with ISOCAM \cite[]{siebenmorgen97}: 
1) single pointing, raster mode, 2) beam switching and 3) spectrophotometry. 
The observations presented in this paper were obtained in raster mode 
where many readouts are put together to build a mosaic of the observed object (called sky image). 
The scanning strategy is made in such a way that each 
readout taken on the sky has some overlap with its neighbours (see Figure~\ref{strategie}).

To illustrate the data reduction procedures, we use two ISOCAM observations of the
gamma-ray burst GRB 970402 first observed by {\em BeppoSAX} \cite[]{feroci97} on April 2.93 UT 1997.
The coordinates (J2000) of the burst were $\alpha = 14^h50^m16^s$, $\delta=-69^\circ19'.9$ with an error
circle of radius 3'. Target-of-opportunity ISO observations (ISOCAM and ISOPHOT) 
were requested by \cite{castro-tirado98}
to detect a transient infrared emission of this burst. Unfortunately the observations did not show such
emission. Nevertheless these particular observations are of great interest for us since the same
field has been observed twice, subsequently within the same orbit, in exactly the same configuration 
to look for a decrease in the GRB emission. 
The observational strategy of the observations is sketched in Figure~\ref{strategie} and 
described in Table~\ref{observation_tab}. 
Comparing the result of our processing on both observations (GRB1 and GRB2 in the following)
allows us to validate the method.

\begin{table*}
\begin{center}
\caption{\label{observation_tab} Journal of the two ISOCAM observations of the gamma-ray burst GRB 970402.}
\begin{tabular}{|l|c|c|}\hline
& GRB1 & GRB2 \\ \hline\hline
Date & April 5 1997 & April 5 1997 \\
ISO Revolution & 506 & 506 \\
Time since activation & 2.12 (hrs) & 3.42 (hrs) \\
Filter & LW10 & LW10 \\
Wavelength & 8.5-15.0$\mu$m &  8.5-15.0$\mu$m\\
Right Ascension J2000 (center) & 14h 50m 18.2s & 14h 50m 18.1s\\
Declination J2000 (center) & -69$^\circ$ 19' 57.1'' & -69$^\circ$ 19' 59.3'' \\
Total number of readouts & 907 & 907\\
Number of raster steps & 8 $\times$ 8 & 8 $\times$ 8 \\
Readouts per positions & 12 & 12 \\
Step size (pixels) & 8 & 8 \\
Exposure time (seconds) & 5.04 & 5.04 \\
Pixel size & 6'' & 6''\\
Comment & Long term transient & Good quality\\\hline
\end{tabular}
\end{center}
\end{table*}

%%%%%%%%%%%%%%%%%%%%%%%%%%%%%%%%%%%%%%%%%%%%%%%%%%%%%%%%%%%%%
%%%%%%%%%%%%%%%%%%%%%%%%%%%%%%%%%%%%%%%%%%%%%%%%%%%%%%%%%%%%%

\section{Building the standard low level sky map}

\label{std_data_processing}

\begin{figure}
\hspace{-0.7cm}
\epsfxsize=9cm
\epsfbox{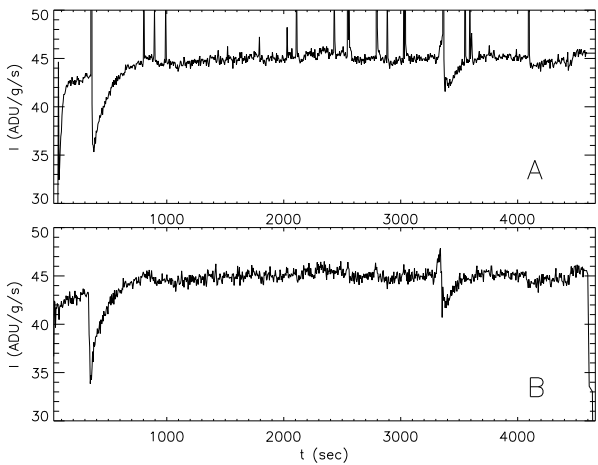}
\caption{\label{deglitch} Temporal evolution of a typical pixel. 
A) Raw data - many energetic electrons have hit the pixel causing instantaneous flux steps
known as fast glitches. A response accident is apparent near $t=350$ seconds caused by an ion impact (slow
glitch).
B) Fast glitches corrected data - short time flux steps have been identified as glitches and 
removed from the data cube.}
\end{figure}

The standard ISOCAM reduction method consists of the following steps:

{\bf Deglitching.}
On any observation, numerous energetic particles impacts
leave traces on one or several pixels of the detector. 
Most of these impacts do not affect the detector response on long time scale 
and are seen as instantaneous flux step ({\em fast} glitches). In Figure~\ref{deglitch}, a typical 
flux history of one pixel as function of time is shown where many fast glitches are apparent. 
Various techniques achieve to suppress these emission spikes \cite[]{starck99}.

{\bf Dark image subtraction.}
The response of the LW detector of ISOCAM is not zero when the detector does 
not receive any photon coming from the sky. 
This dark image must be subtracted from the data.
Temporal variation of the dark image has been observed (\cite{starck99})
and modeled by \cite{biviano98} using a polynomial function of time. 

{\bf Short term transient.}
It has been known since pre-launch measurements \cite[]{perault94} that the ISOCAM detector is affected
by a transient effect, i.e. its response to a given flux step is not instantaneous. 
Recently a new model of the response, based on the physic of Si:Ga arrays \cite[]{fouks95}, 
has been developped by \cite{coulais2000}.
The uncertainty on the corrected flux with this new model is $\sim$ 1\% but
is raises to about 10\% for point sources likely due to charge coupling effects between neighbouring pixels. 
This is not a strong problem here since we are mostly interested in diffuse emission.

{\bf Flat Field.}
Generally, each pixel of a detector does not have exactly the same response
to a given flux.
This spatial variation of the detector response (the {\em flat field} in 
the following) must be taken into account. 
The most straightforward way to compute an image of the flat field is to 
average all readouts of a data cube, and normalize the mean to 1 (\cite{starck99}). 
This technique supposes that all camera pixels have seen the same average flux
along the observation. It is a reasonable approximation in raster
mode where the camera is moving on a region of the sky much larger than the field of view
of the camera, which do not contain any systematic gradient.

{\bf Building the sky image.}
After dividing the data cube by the flat field, the final standard 
data reduction step is to project each readout on the sky to build
the sky image. The LW channel of ISOCAM is affected by a field of view distortion
problem that must be taken into account in this operation
\cite[]{starck99}, especially to co-add properly emission from point sources and small scale structures.

The sky image of the GRB1 observation obtained with the standard data processing 
described in this section is shown in Figure~\ref{mosaic}a.
It is affected by many instrumental problems that prevent to
take advantage of the full sensitivity of the instrument.
In the next section we describe the open problems for the imaging of faint extended
emission.

%%%%%%%%%%%%%%%%%%%%%%%%%%%%%%%%%%%%%%%%%

\section{Open problems}

\label{problems}

\bfi
\hspace{-0.7cm}
\epsfxsize=9cm
\epsfbox{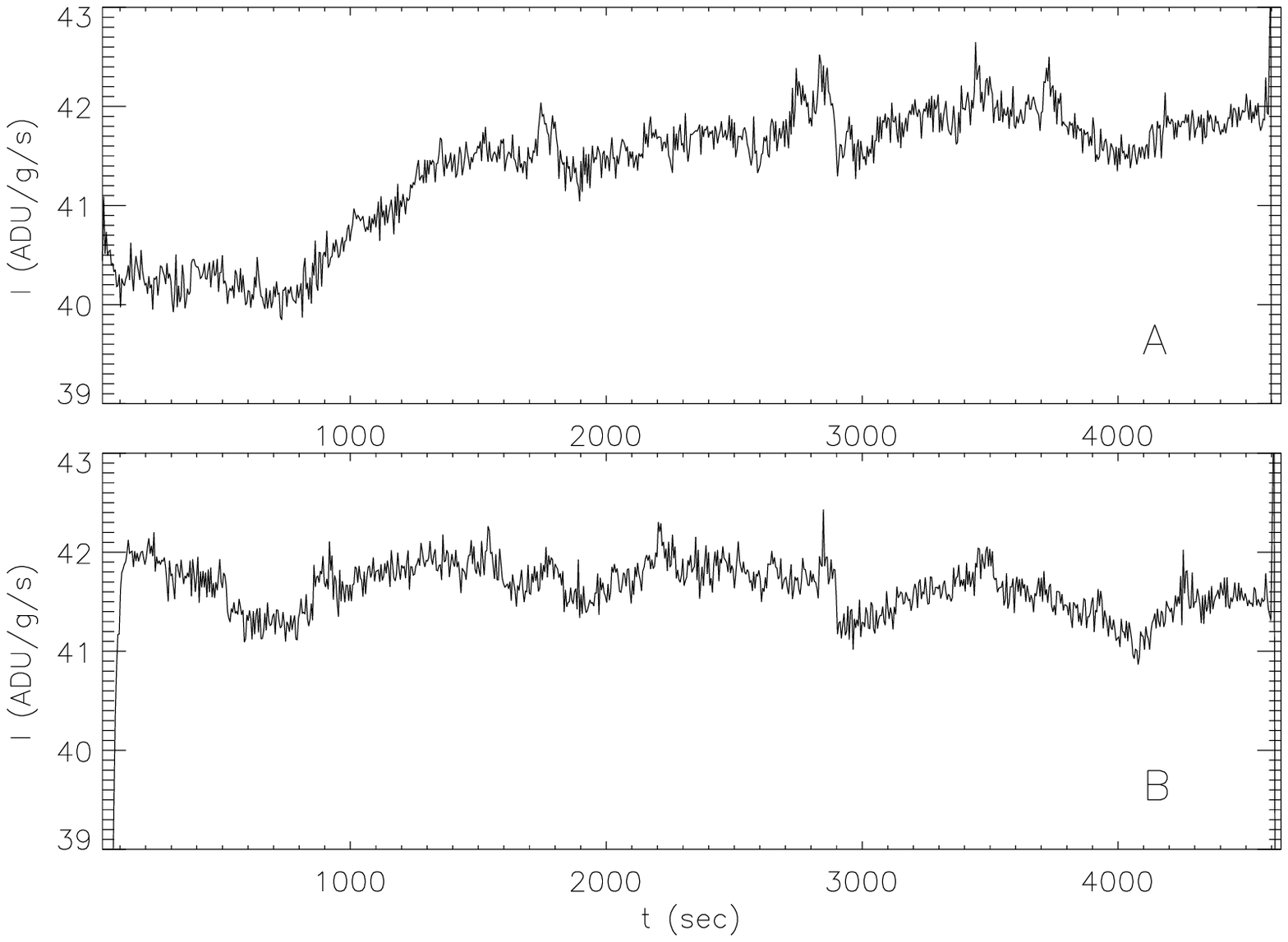}
\caption{\label{stt} Temporal evolution after short term transient correction of the median 
flux observed in the 
$11 \times 11$ central part of the camera, for the GRB1 (A) and GRB2 (B) observations.}
\efi

{\bf Long term transient.} 
Figure~\ref{stt} presents the temporal evolution, 
of the median flux observed in the 
$11 \times 11$ central square of the detector for the GRB1 (a) and GRB2 (b) observations.
One striking feature of Figure~\ref{stt} is the drift in the first observation, not seen
in the second one.
This drift in ISOCAM response is observed quite frequently and is called 
in the following the {\em Long Term Transient} (LTT).
When observed, the amplitude of the LTT is about 5\% 
of the total sky emission like in Figure~\ref{stt}.
One also sees that the first sky image (Figure~\ref{mosaic}a) is completely dominated by the 
long term drift. This is true for any observation affected by a LTT 
of faint extended emission with
contrasts smaller than a few percent of the zodiacal light.

The LTT has been identified during the pre-launch measurements. 
Many calibration observations present a slow and continuous increase of the response after a 
positive flux step.
This rise is gradually attenuating on a variable time scale which can be up to several hours.
Unlike the short term transient, no analytical or physical description 
has been developped up to now to correct this specific effect which
seems to be a general behavior of this type of detector.
\cite{vinokurov91} adressed the problem
of the long term drift in their study of nonlinear response of photoconductor. 
Unfortunatey the analytical approximation they propose does not seem 
to reproduce accurately ISOCAM's LTT.

{\bf Slow glitches.}
Another important problem is related to glitches which modify the response during
a significant time.
This is illustrated in Figure~\ref{deglitch} where 
one notices a significant change in the detector 
response near $t=350$ seconds, just after a glitch impact.
The glitch impact is removed by the standard deglitching algorithm 
but the detector response is significantly disrupted for more than 500 seconds.
In this example the signal is depressed after the glitch, but it can also be enhanced.
These glitches with memory effect (called in the following ``{\em slow glitches}''), 
which affect the response of 
one or more pixels for some time after the impact, are believed to be due to 
heavy ions.
They are responsible for most of the periodic patterns seen in Figure~\ref{mosaic}a.
No model has been developped to correct this instrumental effect.
\cite{desert99} describe the various types of slow glitches known and a method 
to correct the response accident. However, this method 
has been optimised for point sources extraction project where diffuse
emission is not intented to be restored, so we do not use it.

{\bf Ghosts.}
There are also several artificial point sources
(``{\em ghosts}'' in the following) due to 
uncorrected memory effects. After seing a bright point source, the response of a given pixel 
is significantly affected for some time. We have seen in the 
previous section that significant memory effects remain after strong point sources
even after the short term transient correction.
Therefore, as the satellite moves on the sky from one sky position to the other,
pixels that have observed bright point sources are affected by a memory effect, and the 
pattern of each point source 
appears repetitively in the sky image until the response of the pixels gets back to a ``normal'' value.

We conclude that the actual data processing does not adequately take into 
account (1) the long term drift, (2) slow glitches and
(3) short term transients on point sources. The long term drift precludes the study
of large scale emission fainter than 10\% 
of the zodiacal light background. All instrumental effects limit the brightness sensitivity
for small scale structures. Solving these problems is crucial to take advantage of the unique
sensitivity and angular resolution of ISOCAM. This is particularly true for small scale
structures since generally, the contrast of interstellar medium emission decreases 
from large to small scales \cite[]{gautier92}.

\section{New approach for LW-ISOCAM data processing}

\label{new}

\subsection{General description}

\subsubsection{General inversion method}

To address the problems described in \S~\ref{problems} we have developped
a method which uses the fact that a position on the sky has been observed
by several ISOCAM pixels at different times (see Figure~\ref{strategie}).
The redundant information is used to separate the contributions of the sky emission
and of instrumental effects to the observed signal.
This approach has already been used for the IRAS mission (e.g. \cite{okumura91} and \cite{wheelock97})
and may be generalized to every raster type observations, whatever the wavelength of observation.
Formally the processing of astronomical data with spatial redundancy 
could be treated as an inversion problem. 
We can consider that the data observed $O$ is the result of the convolution
of the real sky $S$ with the instrumental function $I$ plus some additive noise $N$:
\begin{equation}
O = I \otimes S + N.
\end{equation}
When the observation has been conducted with spatial redundancy, 
it may be possible to address the data processing problem globally 
by finding $I$, $S$ and $N$ that will reproduce $O$ and minimize the difference between 
measurements obtained at the same sky position.
Such an approach supposes that the main instrumental effects that affect $I$ and $N$
are known enough to constrain the inversion method. Presently this is not the case for ISOCAM.

\subsubsection{Sequential approach}

ISOCAM's response variations are complex 
and presently we are not able to model it with a reasonable number of parameters
which would allow us to address the data processing problem by a global inversion.
We have thus adopted a sequential approach where the instrumental effects are treated
one at a time. Nevertheless, even if we cannot use such a global method,
the idea of minimizing the differences between pixels that have seen the same sky position, 
is the milestone of our approach.
In particular
the ISOCAM long term drift problem is addressed by a least square minimization technique 
based on the fact that, if the detector is slowly reaching stabilization, the measured flux 
is also approaching the real observed flux. Here we suppose that every pixels of the array 
is affected by the same long term drift. The hardness of this assumption will be discussed
in \S~\ref{discussion}.

\begin{figure}
\epsfxsize=9cm
\epsfbox{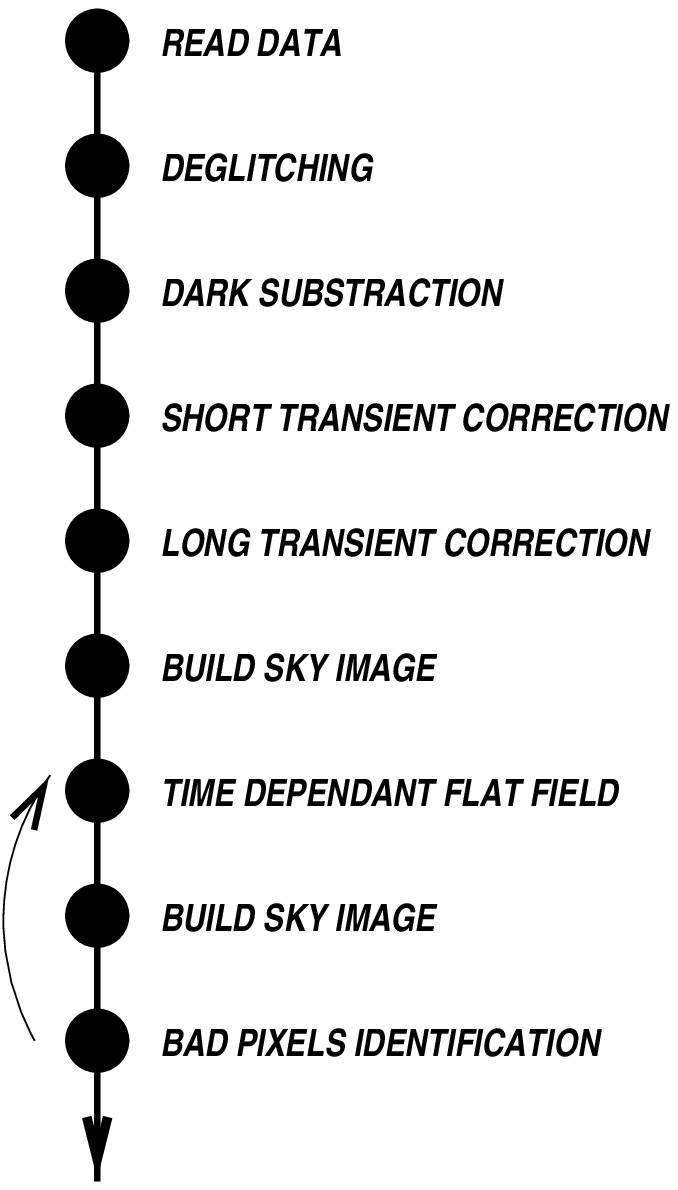}
\caption{\label{pipeline} Data reduction chain illustrating the data processing steps from top to bottom.
	The time dependant flat field and bad pixels identification operations can be done iteratively.}
\end{figure}

Once the LTT is removed, the other response variations (slow glitches, ghosts and
residual transient effects) are corrected by comparing the data cube to the sky image. 
This is done by two operations. First we compute a variable flat field that takes
into account pixel-to-pixel response variations. Second, pixels
that have a flux that departs significantly from the sky image 
are flagged out (these are called {\em bad pixels} in the following). This
last operation removes slow glitches and ghosts. These two operations
(variable flat field and bad pixels identification) can be done several times
to improve the sky image (see Figure~\ref{pipeline}).

\subsection{Long term transient}
\label{LTT}

\subsubsection{Gain or Offset effect ?}

First of all it is important to determine how the long term transient, the flat field 
$F$ and the observed flux $I_{obs}$ are related to the incident flux $I_{sky}$. 
We have considered the three following possibilities for the LTT:
\begin{enumerate}
\item {\em Gain effect}: $I_{obs} = G(t)\,F\,I_{sky}$
\item {\em Offset effect affected by flat}: $I_{obs} = F\,(I_{sky} + \Delta(t))$
\item {\em Pure offset effect}: $I_{obs} = F\,I_{sky} + \Delta(t)$
\end{enumerate}

The two GRB observations are of great help to conclude on the form of the LTT.
As the two observations were done exactly in the same way, it is possible
to subtract directly the two data cubes.
In Figure~\ref{proof_offset}a we show the flux history of two pixels of the GRB2 observation.
As one can see there is no long term drift detectable in this observation. The flux difference
between the two selected pixels is due to a $\sim$25\% flat field difference. 
The difference between observation 1 and 2 for these two pixels is shown in 
Figure~\ref{proof_offset}b. Both pixels
have a very similar drift; the 25\%
difference between the two pixels flux does not appear in figure~\ref{proof_offset}b. 
{\em We then conclude that the LTT does not depend on the signal $FI_{sky}$}.
In this context, the only valid description of the LTT is the pure offset 
effect: $I_{obs} = F\,I_{sky} + \Delta(t)$.
In the following we consider the LTT as a single offset over all the detector;
in other words we suppose that all pixels are affected by the same offset.

\bfi
\epsfxsize=9cm
\epsfbox{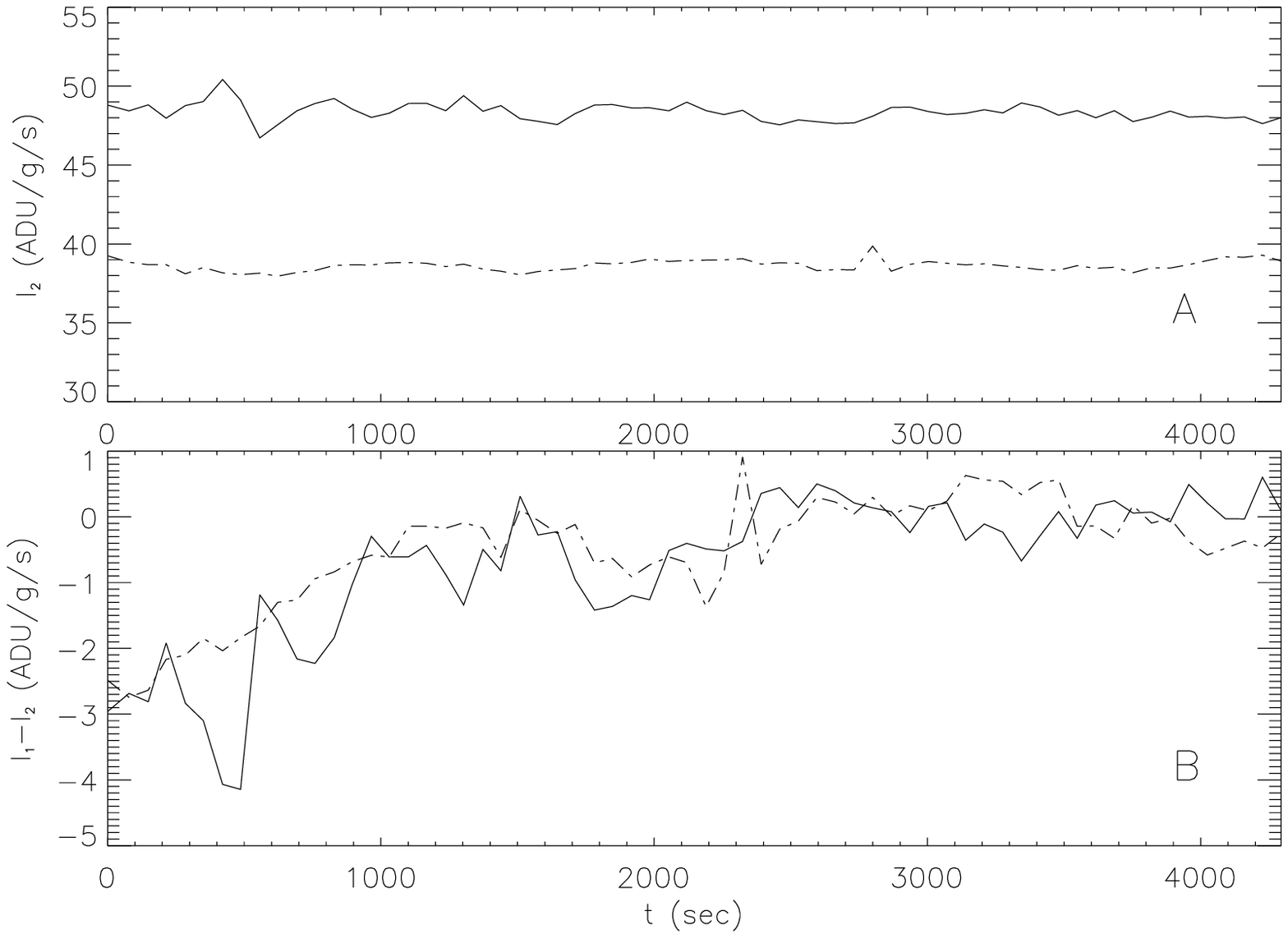}
\caption{\label{proof_offset} (A) Flux history of two pixels of the GRB2 observation. 
This observation is not affected by the LTT. The flux difference between the two pixel is due to
an intrinsic response difference (flat field). (B) Difference between the two GRB observations for
the two pixels considered in (A). Even if the two pixels have very different response, the LTT amplitude
is the same.}
\efi

\subsubsection{Formalism}

\label{formalism}

For a given pixel at a position $(x,y)$ on the detector array and at a given time $t$, 
the observed flux $I_{obs}(x,y,t)$ is related to the temporally varying flat 
field $F(x,y,t)$, the incident flux $I_{sky}(x,y,t)$ and the long term drift $\Delta(t)$ by the following equation: 
\beq
\label{Iobs2Iinc}
I_{obs}(x,y,t) = F(x,y,t)I_{sky}(x,y,t) + \Delta(t).
\eeq
Here we suppose that $\Delta(t)$ is not pixel dependant.

The offset function $\Delta(t)$ is found using equation~\ref{Iobs2Iinc} and the spatial 
redundancy inherent to 
raster mode observations. We determine $\Delta(t)$ by solving a set of linear equations
obtained by comparing flat field corrected intensities of the same sky positions but obtained
at different times.
The flat field $F(x,y,t)$ is computed using the {\em perturbated single flat field}
method (see \S~\ref{second_order}).
The flat field corrected intensities can be written:
\beq
\label{eq2}
\frac{I_{obs}(x,y,t)}{F(x,y,t)} = I_{sky}(x,y,t) + \frac{\Delta(t)}{F(x,y,t)}.
\eeq

In raster mode the camera does not always point at the same position on the sky.
To compare two pixels that have seen the same position on the sky, we must
put all readouts in a common spatial reference frame. Practically we project
$I_{obs}(x,y,t)$ and $F(x,y,t)$ on the plane of the sky (right ascension ($\alpha$) - 
declination ($\delta$) reference frame). Following equation~\ref{eq2}, the difference
between two pixels in that reference frame that have seen the same sky $I_{sky}(\alpha, \delta)$
at different time ($t_i$ and $t_j$) is:
\beq 
\label{difference}
\frac{I_{obs}(\alpha, \delta, t_i)}{F(\alpha, \delta, t_i)} - 
\frac{I_{obs}(\alpha, \delta, t_j)}{F(\alpha, \delta, t_j)} =
\frac{\Delta(t_i)}{F(\alpha, \delta, t_i)} - 
\frac{\Delta(t_j)}{F(\alpha, \delta, t_j)}. 
\eeq
To ensure that we compare fluxes obtained at the same sky position,
the field of view distortion must be accurately taken into account in the projection operation.

The function of interest $\Delta(t)$ is estimated using a least-square minimization technique 
with the following minimization criterion:
\begin{eqnarray}
\label{criterion}
\chi^2 & = & \sum_{\alpha, \delta, t_i, t_j} \left[\frac{I_{obs}(\alpha, \delta, t_i)}{F(\alpha, \delta, t_i)}
 - \frac{I_{obs}(\alpha, \delta, t_j)}{F(\alpha, \delta, t_j)}\right. \nonumber\\
& & - \left.\frac{\Delta(t_i)}{F(\alpha, \delta, t_i)} + \frac{\Delta(t_j)}{F(\alpha, \delta, t_j)} \right]^{2}.
\end{eqnarray}
Here the sum is on all the possible pixel pairs that have seen the same sky.

The function $\Delta(t)$ which minimizes the value of $\chi^2$ is found by solving 
the system determined by:
\begin{equation}
\label{chi2minimum}
\frac{\partial\chi^2}{\partial\Delta(t_i)} =  0.
\end{equation}

Equation~\ref{chi2minimum} represents a standard set of linear equations 
which can be written in a matrix form: $A \Delta(t) = B$, with:
\begin{equation}
A(i,j)_{i\neq j} = \sum_{\alpha, \delta}\frac{1}{F(\alpha, \delta, 
	t_{i})F(\alpha, \delta, t_{j})}
\end{equation}
\begin{equation}
A(i,j)_{i=j} = \sum_{\alpha, \delta, t_{j\cap i}}\frac{-1}{F(\alpha, \delta, 
	t_{i})F(\alpha, \delta, t_{j})}
\end{equation}
\begin{equation}
B(i) = \sum_{\alpha, \delta, t_{j\cap i}} \frac{1}{F(\alpha, \delta, t_{i})} 
\left[\frac{I_{obs}(\alpha, \delta, t_j)}{F(\alpha, \delta, t_j)} - \frac{I_{obs}(\alpha, \delta, t_i)}{F(\alpha, \delta, t_i)}\right].
\end{equation}

The $A(i,j)$ and $B(i)$ terms are computed with all pixels that overlap.
For each pair ($i,j$), the sums are always computed on the positions ($\alpha, \delta$) where
$I_{obs}(\alpha, \delta, t_i)$ and $I_{obs}(\alpha, \delta, t_j)$ are both defined.

Finally $\Delta(t)$ is found from: $\Delta(t) = A^{-1}B$. 
As the second derivative 
of $\chi^2$ is always positive, the solution found for $\Delta(t)$ necessarily minimizes
the $\chi^2$ criterion.
We add an offset to all values of $\Delta(t)$ to force the correction to be zero 
at the end of the observation: $\Delta(t_{end}) = 0$. 
This is justified
by the fact that the long term transient tends to stabilize at the end of an observation.
However, there are several observations not stabilised at the end, so that 
the absolute brightness may be systematically shifted (generally below a few \%).
For the observations of low contrasted clouds on top of a flat large scale emission 
(at least the zodiacal emission), this point is not critical since we always remove the large scale 
emission at the end of our processing.

\subsubsection{Practical implementation}

\label{real_data}

Beside detector noise, two additional sources of uncertainties affect the comparison of 
raster images and thus the 
LTT correction through the minimization algorithm described in the previous section: slow glitches 
and flat field variations along the observation.  
The signal measured on point sources is not identical in the individual readouts,
essentially due to the undersampling of the 
the point spread function for the 6'' pixels. Bright sources are thus an additional source of error but
they can easily be discarded.  The two other noise sources make  
the practical implementation of the formalism a non-straightforward procedure. 

An extensive use of the LTT inversion method presented here on real data has demonstrated 
the extreme importance of an accurate flat field. As the LTT correction is based on the comparison of
the brightness measured by different parts of the array, its result depends on 
the accuracy of the flat field.
In particular, when $F(\alpha,\delta,t)$ is not well estimated, oscillations 
in phase with the rastering of the observations may appear in the correction found.
To pass around this difficulty we compute an approximate LTT correction 
(as oppose to the exact correction which is the solution of equation~\ref{chi2minimum}).
The study of low contrast ISOCAM data affected by long term drifts has shown that, in some cases, 
the LTT can be approximated by the sum of two exponential functions, one upward and one
downward:
\begin{equation}
\Delta(t) \approx P \times \exp{\left(- Q \times t^R\right)} - S \times \exp{\left(- T \times t^U\right)}
\end{equation}
where $P$, $Q$, $R$, $S$, $T$ and $U$ are strictly positive.
By approximating the LTT in that way, we greatly simplify the problem as $\Delta(t)$ depends now
only on six parameters.
As for the exact solution, the approximated solution is
found by minimizing the criterion of equation~\ref{criterion}, using an IDL curve 
fitting program (MPFIT). The obvious advantage
of this method is that the approximated LTT correction is smooth, getting rid of
the oscillations found in some exact solutions. 
%Nevertheless, on very contrasted observations, 
%where the flat field is more difficult to estimate accurately,
%this approximated solution may also be unrealistic. 

{\bf  On the other hand, we have observed cases where this approximation 
of the LTT does not hold. Sometimes, the LTT shows some oscillations that can dominate the
emission in low contrasted fields (e.g. cirrus clouds - see \cite{miville-deschenes2000}). 
In these cases, an accurate estimation
of the flat field and the use of the exact solution for the LTT are mandatory. }

\subsubsection{LTT correction for the GRB1 observation}

Figure~\ref{ltt_grb} presents the LTT corrections found for the GRB1 observation. 
A {\em perturbated single flat field} (see \S~\ref{second_order}) was used and point 
sources were discarded to compute the exact and approximated corrections. 
The approximated LTT correction is smoother than the exact one, which oscillates 
with a $\sim 0.2$ ADU/g/s amplitude and a period of $\sim 1800$ seconds. 
{\bf In this case we have applied the approximated LTT correction.}
The sky image computed after that correction has been done is presented in Figure~\ref{mosaic}b. 
One sees that it is mandatory to apply the LTT correction since its amplitude (3 ADU/g/s) is nearly
three times the amplitude of the emission of interest (1.1 ADU/g/s).
At this stage we have used a single flat field to compute the sky image. It is necessary
to use a variable flat field to correct the artefacts (e.g. periodic patterns) seen in Figure~\ref{mosaic}b.

\bfi
\epsfxsize=9cm
\epsfbox{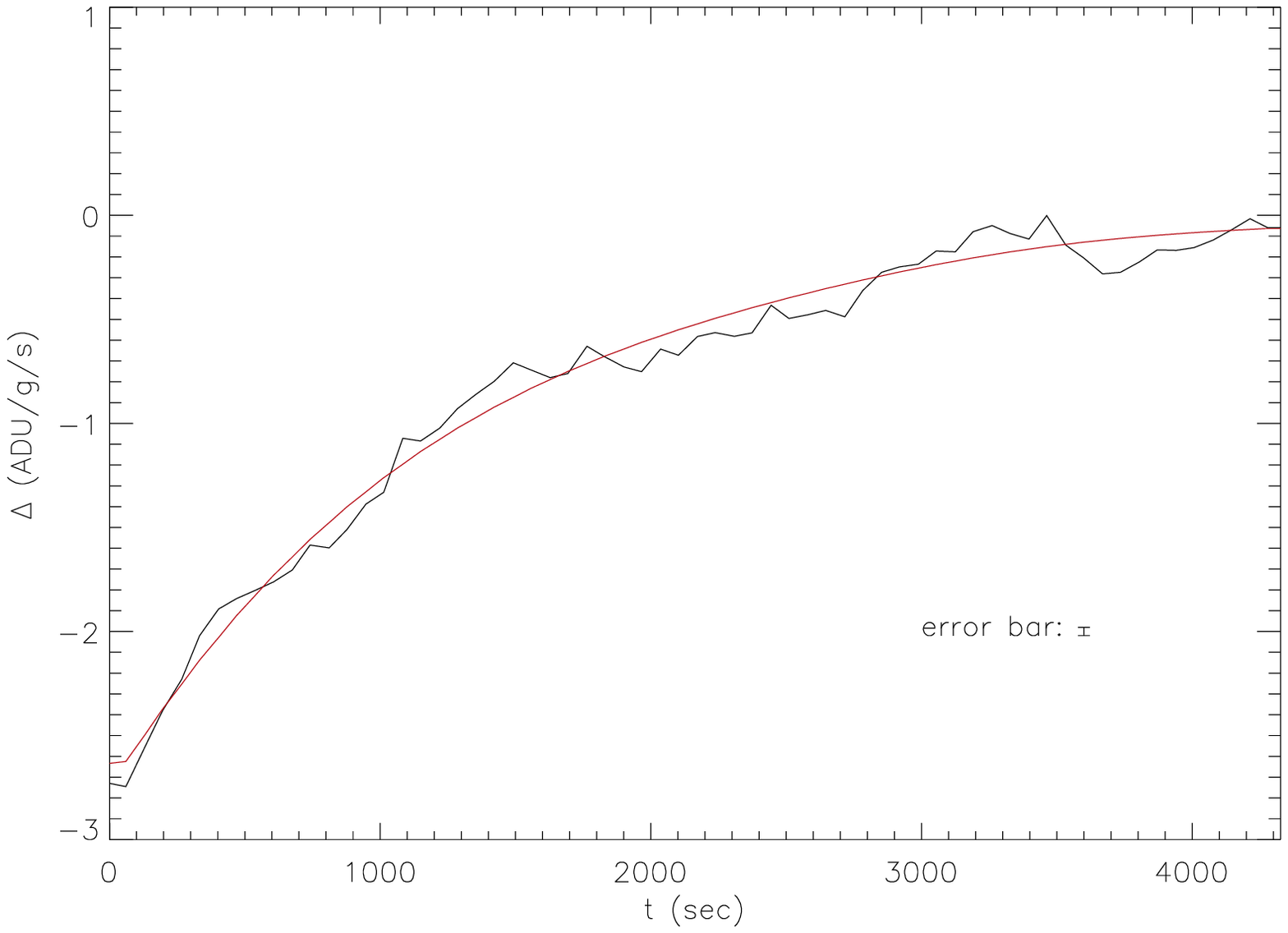}
\caption{\label{ltt_grb} Exact and approximated LTT correction for the GRB1 observation.
Estimating the error on the LTT correction is difficult as it is very sensitive
to the accuracy of the flat field used.
The error bar represented here is the statistical error (0.08 ADU/g/s - see \ref{error_ltt}).  
A more realistic error determination would be related 
to the accuracy of the flat field. For the GRB observations, a 1\% uncertainty on the flat field 
corresponds to a 0.4 ADU/g/s error on the LTT correction.}
\efi

\subsection{Variable Flat Field}
\label{variable_flat}

Now that the LTT has been corrected, we take into account the pixel-to-pixel 
temporal variations of the detector response. These response variations are observed
at various timescales. At short timescales they are due to bad short term transient
correction, particularly on point sources. On longer time scales they are mainly 
caused by slow glitches (see \S~\ref{problems}).
The use of a single flat field (see \S~\ref{std_data_processing})
does not take into account these temporal variations (that represent  $\sim$1-3 \% of the average flat field)
preventing to benefit of the optimal ISOCAM sensitivity. 
To go further in the data processing, we try to correct these pixel-to-pixel response
variations with a time-dependant flat-field
(called ``{\it variable flat field}'' in the following).

For LTT corrected data,  
the observed flux $I_{obs}$ at position $(x,y)$ on the array and at time $t$ is
\beq
\label{Iobs_gen_equ}
I_{obs}(x,y,t) = F(x,y,t)\, I_{sky}(x,y,t),
\eeq
where $F(x,y,t)$ is the instantaneous response of pixel $(x,y)$ at time $t$. 
Thus flat field and sky structures are mixed together in $I_{obs}(x,y,t)$. 
In the following we show how the flat field variations can be extracted from the 
data by estimating $I_{sky}(x,y,t)$ and by taking advantage of the spatial redundancy.

\label{sky_sub}

In raster observation mode, many pixels of the data cube have seen the same position 
($\alpha, \delta$) on the sky. By averaging all these pixels we reduce the noise due to 
instrumental effects on the computation of the sky image $I_{sky}(\alpha, \delta)$. 
The estimate of $I_{sky}(x,y,t)$ is made by an inverse projection of $I_{sky}(\alpha, \delta)$ 
on each readout of the data cube. Then $F(x,y,t)$ is computed by averaging \linebreak
$I_{obs}(x,y,t)/I_{sky}(x,y,t)$ 
with a sliding window on the time axis (see \S~\ref{sliding_mean}).
Here are the guidelines of this method:
\begin{enumerate}
\item Construct a sky image.
\item Smooth (median smoothing) the sky image with a $10\times10$ window.
\item Compute an {\em ideal cube} $I_{sky}(x,y,t)$ 
	by projecting the smoothted sky image on each readout of the data cube.
\item Smooth (median smoothing) $I_{obs}(x,y,t)/I_{sky}(x,y,t)$ on the time axis. The size of the
	smoothing window should be of the order of the time spend on 5 different sky positions.
\end{enumerate}

\begin{figure*}
\hspace{2cm}
\epsfxsize=18cm
\epsfbox{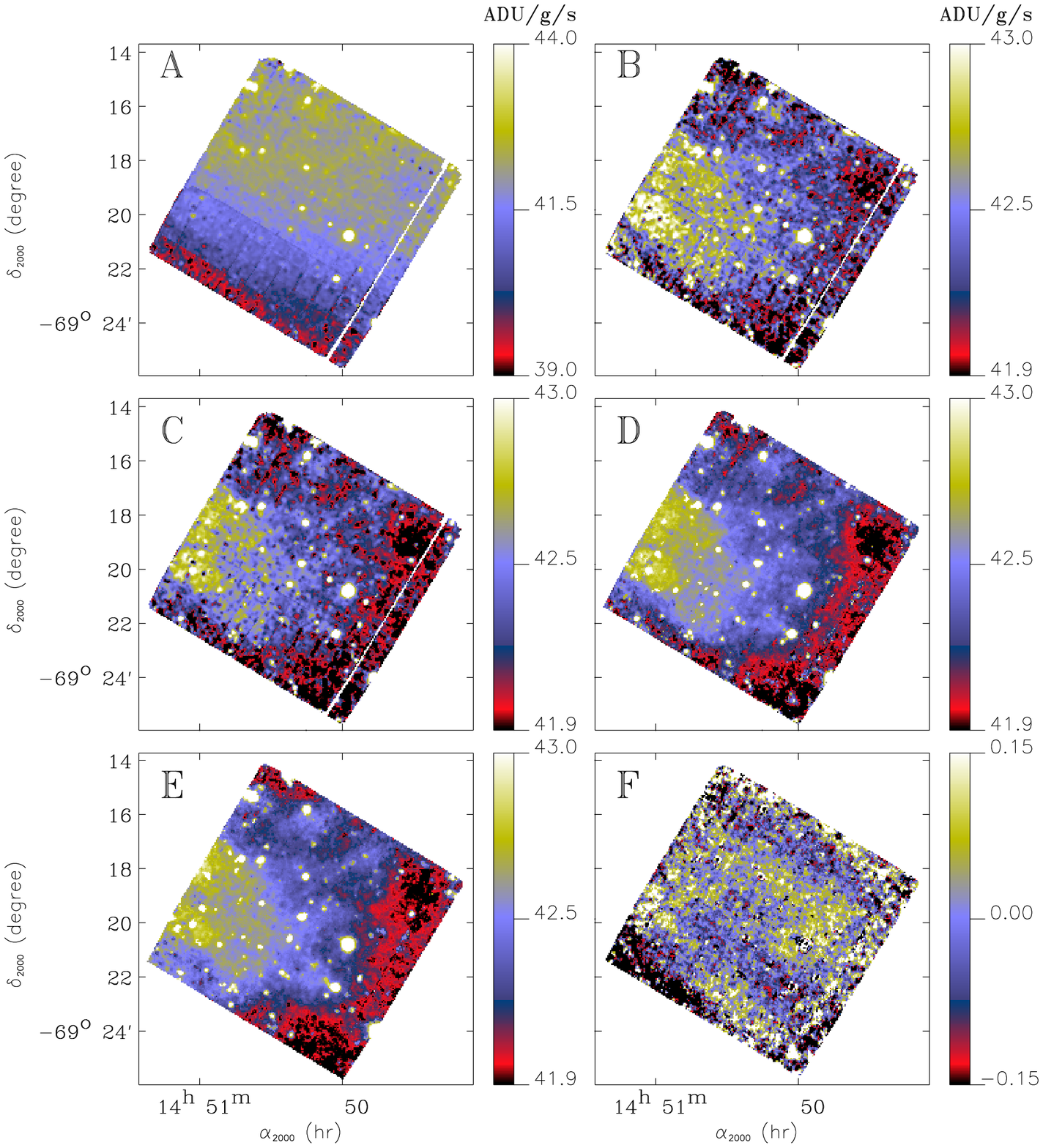}
\caption{\label{mosaic} LW10 images of a the first GRB970402 observation
after deglitching, dark substraction, transient correction (A), long term transient correction (B),
variable flat field (C) and bad pixels removal (D). Image (E) is the final map of the second GRB970402 observation
and image (F) is the difference between (D) and (E). For these observations, 1 ADU/g/s corresponds to 0.242 mJy/pix
or 0.286 MJy/sr.}
\end{figure*}

The sky image of the GRB1 observation, obtained with the variable flat field,
is shown in Figure~\ref{mosaic}c. The size of the filtering window on the time axis
is 100 (corresponding to $\sim$500 seconds). 
As seen in Figure~\ref{mosaic}c, the variable flat field removes almost all 
periodic patterns due to high frequency variations of the detector response. 
Compare to other methods we have tested (see~\ref{sliding_mean} and \ref{second_order}),
this variable flat field gives by far the best results. 

\begin{figure}
\epsfxsize=9cm
\epsfbox{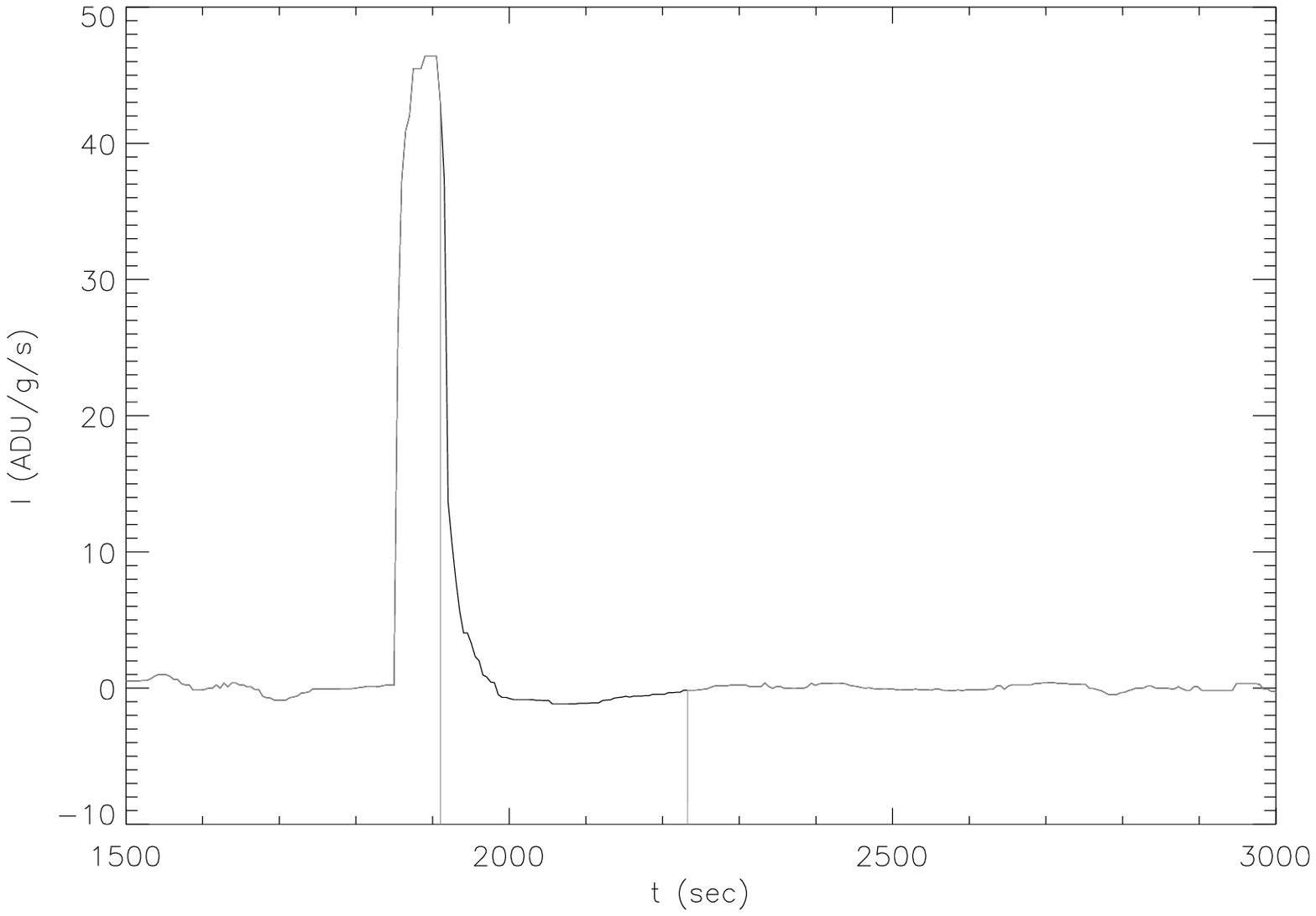}
\caption{\label{ghost_identification} Ghost identification. Flux history of a pixel which has observed
a point source. The flux values after the point source are affected by a memory effect and are identified as a ghost
(flagged out).}
\end{figure}

\subsection{Bad pixels identification}

The deglitching process used at the beginning of the reduction chain (see \S~\ref{std_data_processing}) 
removes extremely deviant pixels that have been hit by cosmic rays. 
To go further in the noise minimization process, we take again advantage of the spatial
redundancy.

\subsubsection{Ghost removal}

Memory effects often appear after a strong flux step (e.g after a point source)
due to improper short term transient correction; this is what we call ghosts.
To identify pixels affected by such effects, we look on the flux history of every pixel for
memory effect after a flux step (see Figure~\ref{ghost_identification}). 
Again we use the redundancy information to improve the identification
of ghosts. Here is how we proceed:

\begin{enumerate}
\item Smooth (median smoothing) the sky image with a 3 $\times$ 3 window.
\item Compute an {\em ideal cube} by projecting the smoothted sky image on each readout of the data cube.
\item Compute the residual cube = $|$data cube - {\em ideal cube}$|$.
\item Identify strong flux steps as residual flux above 10 $\times$ the noise level.
\item On the time history of each pixels, identify ghost as residual fluxes above or under the noise level
	{\em after a strong flux steps}.
\item Flag ghosts in the original data cube.
\end{enumerate}

\subsubsection{Noise reducer}

We can go further in the reduction of the noise level by working sky position by sky position instead of
working on the time history of every pixel. The idea is to look at pixels in the data cube that have
seen the same sky position and discard deviant flux values.

First the sky image is smoothed (median smoothing) with a 10 $\times$ 10 window. 
Then, for a given sky position ($\alpha$, $\delta$), we compare the $N$ pixels in the data cube 
that have seen that position with the flux of the smoothed sky image. 
For most of the sky position, the $N$ pixels are distributed
around the smoothed sky estimate. On the other hand, at point sources positions, the $N$ pixels will
be generally above the smoothed sky estimate. Furthermore, it is also possible to find sky positions where most
of the $N$ pixels have fluxes under the smoothed sky level. Here is how we deal with each case:
\begin{enumerate}
\item {\em Most of the $N$ pixels are around the smoothed sky level ($SML$).}\\
	The new sky flux is the average of ($SML - 3 \times noise < I < SML + 3 \times noise$)
\item {\em Most of the $N$ pixels are above the smoothed sky level.}\\
	The new sky flux is the average of $I > SML$
\item {\em Most of the $N$ pixels are under the smoothed sky level.}\\
	The new sky flux is the average of ($SML - 3 \times noise < I < SML$).
\end{enumerate}
This method allows to globally reduce the noise level and keep a good photometry
of point sources. This is the final data processing step.
The sky images obtained after bad pixels removal of the first and second GRB observations 
are presented in Figure~\ref{mosaic}~D and Figure~\ref{mosaic}~E. 

\section{Assesment of the method}

\label{discussion}

\subsection{Comparison of the two GRB observations}

\label{comparison}

From the comparison of the final maps of the two GRB observations (Figure~\ref{mosaic}~D and Figure~\ref{mosaic}~E)
we can estimate the reliability of our processing. At first glance we see that the structure
of the diffuse emission is very similar in both maps; the LTT correction
applied seems to restore properly the large scale structure.
Furthermore, almost all point like structures are present in both maps giving confidence
in our bad pixels identification.

The difference of the two final sky images is shown in Figure~\ref{mosaic}F.
It is dominated by small scale structure noise but large scale structures are also apparent.
These are probably due to error in the LTT correction.
Extra noise is seen at point source positions. This was expected as memory effects
are not fully corrected on point sources and that we are undersampling the PSF. 
One also notices that the noise level is higher at the edges of
the difference map, due to less redundancy in these regions.
The standard deviation of the difference map (Figure~\ref{mosaic}F) in the central part
of the field is 0.06 ADU/g/s.
Therefore, the noise on each GRB final maps can be estimated at $0.06/\sqrt{2} = 0.04$ ADU/g/s.

\begin{figure}
\hspace{2cm}
\epsfxsize=9cm
\epsfbox{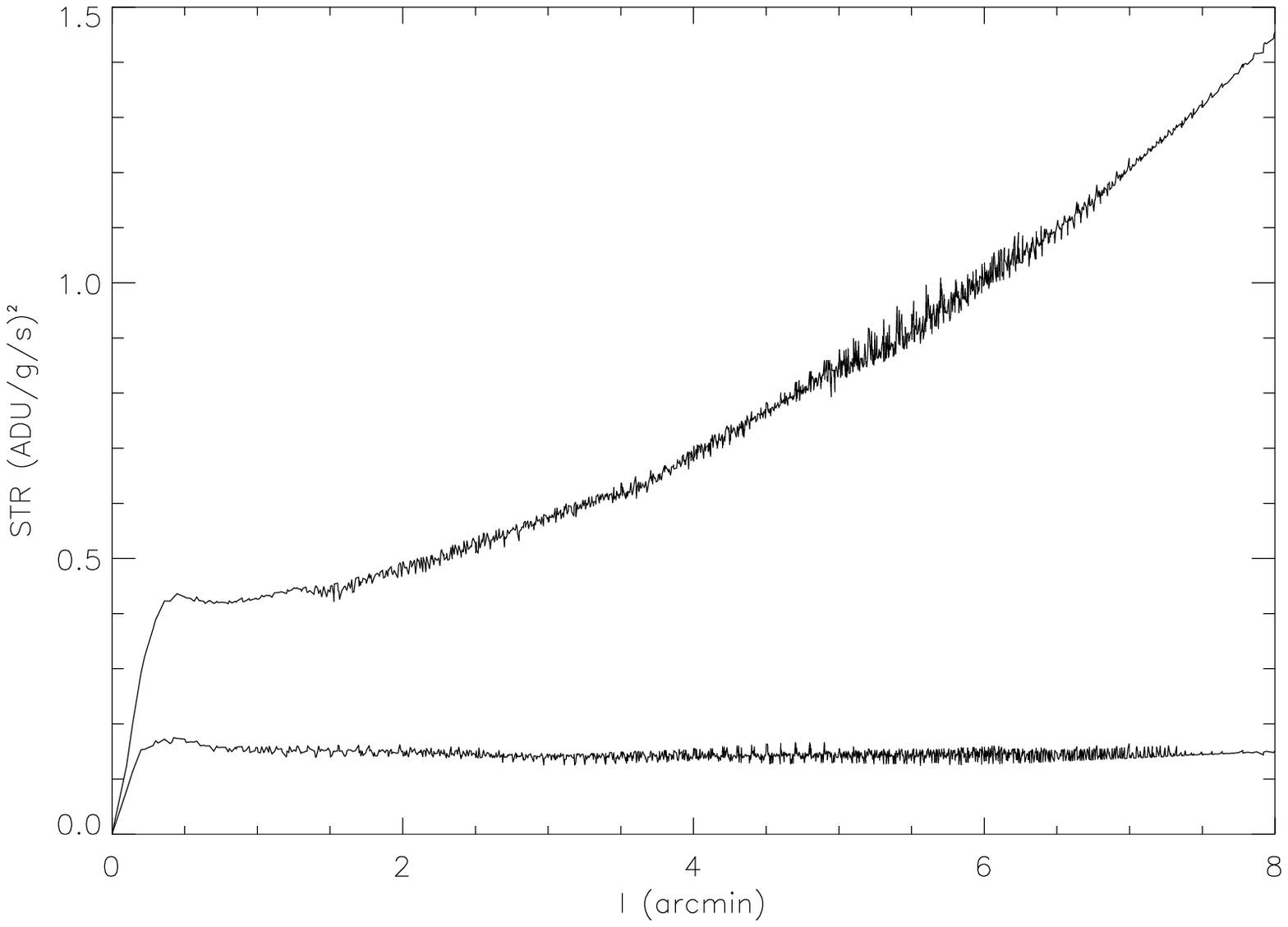}
\caption{\label{str_diff_grb} Structure function of the difference between the two
GRB final maps (bottom line) and of the difference between the first map of GRB1
and the final map of GRB2 (top line).}
\end{figure}

It is clear from the difference map (Figure~\ref{mosaic}F) and from the impact of the LTT
on the sky image that noise is present at all scales. The processing presented in this paper
affects the signal at various scales. To characterize the noise as a function of angular scale
we use the structure function of second order:
\begin{equation}
B^2({l}) \equiv \frac{\sum [I_{sky}({r}) - I_{sky}({r}+{l})]^2}
{N({l})},
\end{equation}
where the sum is over all the $N(l)$ pairs of points
separated by a distance $l$.
To estimate the noise at each scale in the first map, the one affected
by all instrumental effects, we compute the structure function on the difference
between the first map of the GRB1 observation and the final one of the GRB2 observation. 
This difference map, where the sky is removed, is dominated by the noise of the GRB1 observation. 
The structure function of this difference
map rises strongly from small to large scale (see Figure~\ref{str_diff_grb}),
mainly due to the presence of the LTT. To estimate the noise at the end of the processing,
we have computed the structure function on the difference between the final maps of the two GRB observations
(see figure~\ref{mosaic}F). This time the structure function is very flat
(see figure~\ref{str_diff_grb}). The noise level is reduced by a factor ten at 8 arcmin scale
and a factor two at the resolution limit.
This indicates that the noise level has been lowered at all scales and that it
is now uniform at all scales.

\subsection{Study of the noise sources}

\begin{figure*}
\hspace{2cm}
\epsfxsize=18cm
\epsfbox{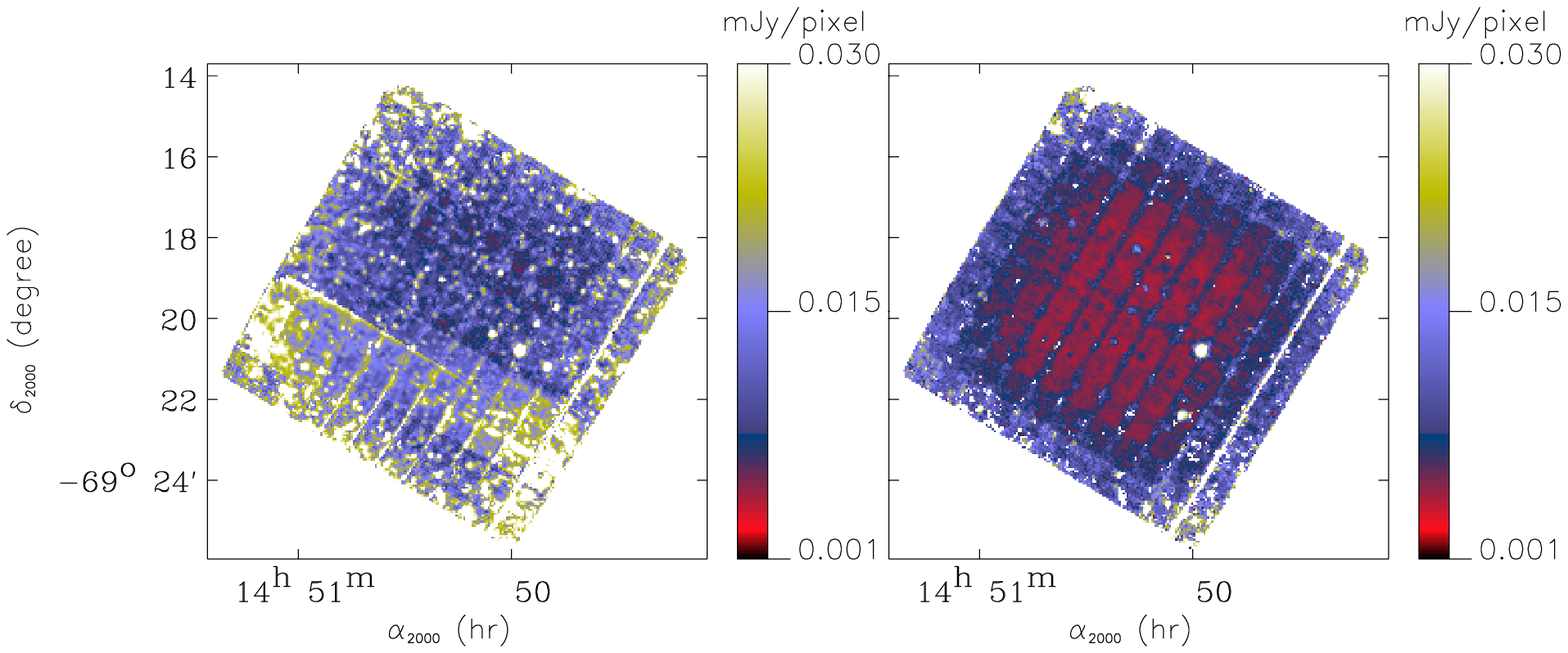}
\caption{\label{error_map} Noise maps before (left) and after (right) the processing.
For this observation 1 ADU/g/s corresponds to 0.242 mJy/pix.}
\end{figure*}

\begin{table*}
\begin{center}
\caption{\label{noise} Noise table. The upper panel gives the noise level computed on the
flux history of a pixel. The bottom one indicates the noise level in the sky image at each step
of the processing.}
\begin{tabular}{|l||l|c|c|c|c|c|}\hline
\multicolumn{7}{|c|}{Noise on a single ISOCAM readout (readout and photon noise)}\\\hline
\multicolumn{6}{|l|}{Ground alibration noise estimate (for $I=41.5$ ADU/g/s)} & 1.15\\
\multicolumn{6}{|l|}{Noise measured on the flux history of a pixel (GRB1)} & 1.15\\\hline\hline
     	& Sky Image     	& Figure& \multicolumn{2}{c|}{$\delta^a$} &N$^b$	&$\sigma$\\ \cline{4-5}
     	&               	&	& (ADU/g/s) 	& ($\mu Jy$)	&  		& (ADU) \\ \hline
GRB1 	& Before LTT correction	& 8A	& 0.062		& 15.0		& 76 		& 2.72$^c$ \\
	& After LTT correction	& 8B	& 0.053		& 15.8		& 76		& 2.32$^c$ \\
	& Variable Flat Field	& 8C	& 0.046		& 11.13		& 76		& 2.01$^c$ \\
	& Final map		& 8D	& 0.031 	& 7.50		& 58		& 1.21$^c$ \\ \hline
GRB2	& Final map		& 8E	& 0.030		& 7.26 		& 60		& 1.15$^c$ \\ \hline
\end{tabular}\\
$^a$ median error value of each sky image. \\
$^b$ median redundancy of each sky image. \\
$^c$ $\sigma$ = Error $\times \sqrt{\mbox{Redundancy}} \times$ Integration time.\\
\end{center}
\end{table*}

The goal of this section is to show that the high spatial frequency noise of our maps
is close to the optimal value obtained with stabilized ground calibration data with no glitches.
The data are affected by many sources of noise. First, there are the classical
quantum photon noise and the detector readout noise which have been
extensively studied in the pre-launch calibration phase \cite[]{perault94}.
A conservative value of the readout noise is given in the ISOCAM cookbook: 1.5 ADU/g. 
Secondly, memory effects (short term transient, long term transient 
and slow glitches) and fast glitches with small amplitude increase substantially the noise levels. 
These non-gaussian events may prevent to reach the optimal sensitivity.

The noise level is measured on the flux history of pixels for which
fast glitches have been removed. We have selected pixels not affected by slow glitches.
We quantify the noise using the standard deviation of the high frequency component of the pixel flux history.
We see on Table~\ref{noise} that the noise level of the two GRB observations is in total 
agreement with the readout and photon noise estimated from calibration data (for a 41.5 ADU/g/s flux). 

The flux $I_{sky}$ at a given position in the final sky image is the average of $N$ independant
flux measurements. The error $\delta$ on $I_{sky}$ can be estimated by
\beq
\delta = \frac{\sigma}{\sqrt{N}},
\eeq
where $\sigma$ is the standard deviation of the $N$ flux measurements used to compute $I_{sky}$.
We have computed the error map at each step of the processing. 
In Figure~\ref{error_map} is shown the error $\delta$ of each sky position for the
sky image obtained before the LTT correction (Figure~\ref{error_map}A) and 
for the final sky image (Figure~\ref{error_map}B) of the GRB1 observation. 
In the Figure~\ref{error_map}A one sees an enhancement of the error in the southern part
of the image, due to the presence of the LTT. Glitches and periodic flat defects appear as 
noise peaks in this error map. The structure of the final error map (Figure~\ref{error_map})
is dominated by the redundancy effect: the noise is higher on the edges of the map
as less flux measurements were obtained in these regions. 

For each sky image of Figure~\ref{mosaic}, Table~\ref{noise} lists
the median error $\delta$, the median redundancy $N$ and the median standard deviation of the $N$ 
flux measurements averaged at each sky position. The first thing to notice is that the noise level
decreases gradually through the processing. 
In the final sky image, the median $\sigma$ is only 5\%
above the noise calibration measurement for the GRB1 observation. The noise in the 
GRB2 observation is exactly the one obtained with stabilized ground calibration data with no glitches. 

The dispersion of the difference between the two final maps (divided by $\sqrt{2}$) is 0.04 ADU/g/s (see \S~\ref{comparison}) 
which is 35\% above the noise level computed on each final map (0.03 ADU/g/s - see Table~\ref{noise}).
This 35\% difference is partly due to the increase noise on point sources
and to the imperfection of the LTT correction. Nevertheless, considering 
the amplitude of the instrumental effects (3 ADU/g/s for the LTT) we think
that the comparison between the two GRB observations 
is a strong validation of the whole processing.

Finally we conclude from the numbers of Table~\ref{noise} 
that the noise level in our final maps are dominated by the readout and photon noise. 
The other instrumental effects are corrected in the data reduction. 
Such ISOCAM sensitivity has been obtained in recent point sources extraction studies \cite[]{desert99,aussel99}
where the low frequency diffuse emission is removed. 
It is the first time that such a sensitivity is reached for the emission structure on all scales.

\section{Conclusion}

\label{conclusion}

The data processing techniques presented here are based on the fact
that each position on the sky has been observed several times and by different
pixels along the observation.
Taking advantage of this redundancy information inherent to raster mode observations,
we are able to control some instrumental effects which degrade the signal, reduce the noise level
and finally bring images obtained with the LW channel of ISOCAM to a quality level close to optimal. 
The principal of the method presented could be generalized to all observations with spatial redundancy.

The quality of the final images is limited by our ability at separating the various instrumental effects.
We have adopted a sequencial approach where each of the problems is treated one after the other,
but a global minimization would be a better way to find the optimal solution. However such
an approach seems to be out of reach for ISOCAM since several instrumental problems 
(e.g. the LTT and response variation after glitches and point sources) are presently not understood.

All raster ISOCAM observations can benefit from the
data processing described here. This includes mapping of solar system extended objects (comet dust trails), 
nearby clouds and star forming regions, images from diffuse emission in the Galactic plane
and external galaxies. Several publications based on the interpretation of ISOCAM
observations reduced by the present method are being prepared.

\acknowledgement{We thank the different teams at IAS, CEA-Saclay, ENS-Paris and ESA for their outstanding 
work and continuous support during all phases of the ISO project. The Fond FCAR du Qu\'ebec and the National 
Science and Engineering Research Council of Canada provided funds to support this research project. 
MAMD warmly thanks Leo Metcalfe for very helpfull discussions.}

\begin{appendix}

\section{Flat fielding}

\subsection{Averaging on a sliding window}
\label{sliding_mean}

One way to estimate a flat field for each readout of a data cube consists 
to compute a sliding window average of the data cube:
\begin{equation}
F(x,y,t_i) = \frac{1}{N}\sum_{t_j=t_i-N/2}^{t_i+N/2} I_{obs}(x,y,t_j),
\end{equation}
where $N$ is the size of the sliding window.

The variable flat field is normalized to 1 over the 11$\times$11 central
part of $F(x,y,t)$:
\begin{equation}
F(x,y,t_i) = \frac{F(x,y,t_i)}{\left< F(10:21,10:21,t_i) \right>}.
\end{equation}

Here we suppose that during the $N$ readouts the camera has observed many positions on the sky 
and each pixel of the detector has seen the same flux in average.
This is a good approximation if the sky observed is rather smooth and does not contain
too much small scale structures.
If the emission is fairly uniform on the detector, 
the number of images required to compute the flat field may be relatively 
small. The choice of $N$ depends on the number of readouts taken per sky position
and on the contrast of small scale structures
(in our case we use $N=100$). 
Extreme high and low fluxes pixels were rejected (top and bottom 15\%)
in order to exclude glitches and point sources.
The main limitation of this method is that sky structures may be still present in the variable flat field.

\subsection{Perturbated single flat field}

\label{second_order}

The {\em perturbated single flat field} method is based on the following assumption. We suppose
that the variable flat field is, to first order, a single flat field $F(x,y)$ 
computed on the whole data cube 
(see \S~\ref{std_data_processing}). The temporal variations of the flat field are treated as perturbations
of the single flat, and we have:
\begin{equation}
I_{obs}(x,y,t) = [1 + \delta(x,y,t)]F(x,y)I_{sky}(x,y,t),
\end{equation}
where $\delta(x,y,t)$ is the perturbation term that take into account flat field deformations.

Flat deformations are mainly due to slow glitches and bad short term transient correction (especially on point sources).
Therefore, for one given readout, they are considered as small scale defects, and $\delta(x,y,t)$
is dominated by high frequency structures in the $(x,y)$ space.

On large scale, the quantity  
\begin{equation}
\frac{I_{obs}(x,y,t)}{F(x,y)} = I_{sky}(1 + \delta(x,y,t))
\end{equation}
is dominated by real large scale structures ($I_{sky}(x,y,t)$) and not by flat defects ($\delta(x,y,t)$).

The low frequency sky emission $I_{SLF}(x,y,t)$ is estimated by smoothing (median filtering) 
each readout of \linebreak
$I_{obs}(x,y,t)/Flat(x,y)$. The size of the smoothing window has to be smaller
than the smallest sky structures and larger than the largest flat defects. In most cases
a compromise must be found (typically a 7$\times$7 window). Then the flat field deformations are estimated using:
\begin{equation}
\label{second_order_eq}
(1+\delta(x,y,t)) \simeq \frac{I_{obs}(x,y,t)}{F(x,y)I_{SLF}(x,y,t)}.
\end{equation}
We get rid of residual high frequency real sky structures by applying a temporal sliding
average (see \S~\ref{sliding_mean}) over the right hand side term of equation~\ref{second_order_eq}. 

The variable flat field is finally obtained with:
\begin{equation}
Flat(x,y,t) = [1 + \delta(x,y,t)]F(x,y)
\end{equation}
The {\em averaging on a sliding window} method (see \S~\ref{sliding_mean}) follows low frequency temporal flat field 
deformations but, as a limited
number of readouts are averaged together, sky structures are still present in the flat field.
The {\em perturbated single flat field} is a better estimate since low frequencies of the incident flux cube are removed.
%However $I_{sky}$ is still not perfectly estimated (median smoothing). In the next section we present a method where the sky is
%totally removed from the data cube to compute the variable flat field.

\section{LTT error determination}
\label{error_ltt}
The error on the LTT correction is estimated using the standard deviation
constructed with all the pixel pairs used to compute $\Delta(t)$.
The offset correction is based on Equation~\ref{difference}. For a pixel pair 
(($\alpha, \delta, t_i$) and ($\alpha, \delta, t_j$)) that have seen the same sky at different time, 
the offset at time $t_i$ is given by:
\begin{equation}
\begin{array}{ll}
\Phi(\alpha, \delta, t_i, t_j) = & 
F(\alpha, \delta, t_i) \times\\
& \left[ \frac{I_{obs}(\alpha, \delta, t_i)}{F(\alpha, \delta, t_i)}
 - \frac{I_{obs}(\alpha, \delta, t_j)}{F(\alpha, \delta, t_j)} + 
	\frac{\Delta(t_j)}{F(\alpha, \delta, t_j)} \right]. 
\end{array}
\end{equation}

We have evaluated the statistical error on the offset correction $\Delta(t_i)$:
\beq
\sigma(t_i) = \sqrt{\frac{1}{n(n-1)} 
	\sum_{\alpha, \delta, t_{j\cap i}}(\Delta(t_i) - \Phi(\alpha, \delta, t_i, t_j))^2}
\eeq
where $n$ is the number of pixel pairs.
%We have checked with simulations that this statistical error is a realistic estimate.

\end{appendix}

\end{document}